\documentclass[prb,preprint,superscriptaddress]{revtex4}

\usepackage{graphicx}
\usepackage{epsfig}
\usepackage{fancyhdr}

\def\PRL{\textit{Physical Review Letters}}
\def\PRB{\textit{Physical Review B}}
\def\APL{\textit{Applied Physics Letters}}
\def\JAP{\textit{Journal of Applied Physics}}

\begin{document}
\preprint{{\it Am. J. Phys.}, {\bf 77}(12), 1112-1117 (2009).}

\title{A student's guide to searching the literature using online databases}

\author{Casey~W.~Miller}
\email{cmiller@cas.usf.edu}
\affiliation{Department of Physics, University of South Florida,
4202 E. Fowler Avenue, Tampa, Florida 33620 USA}

\author{Michelle~D.~Chabot}
\email{mchabot@cas.usf.edu}
\affiliation{Department of Physics, University of South Florida,
4202 E. Fowler Avenue, Tampa, Florida 33620 USA}

\author{Troy C. Messina}
\email{tmessina@centenary.edu}
\affiliation{Department of Physics, Centenary College of Louisiana, 2911 Centenary Blvd., Shreveport, Louisiana 71104 USA}

\begin{abstract}
A method is described to empower students to efficiently perform general and literature searches using online resources. The method was tested on undergraduate and graduate students with varying backgrounds with scientific literature. Students involved in this study showed marked improvement in their awareness of how and where to find accurate scientific information.
\end{abstract}

\maketitle

\section{Introduction}
One of the most important tools for researchers is the ability to find and judge the work of other scientists. These talents are developed over time, but can be expedited by a working knowledge of how to efficiently use internet databases. Literature search tutorials provided by libraries\cite{UCBtutorial, Manchestertutorial} are too general for students to easily apply to specific databases or disciplines, causing novice researchers to spend large amounts of time performing what are often fruitless searches. To assist students some institutions have implemented courses devoted to science literature searches. A recent study focusing on the health sciences concluded that even university faculty members are unaware of the common tools available for literature searches.\cite{healthsci}

In this article we demonstrate how students can find the most influential papers on a general topic, and then find the most influential papers related to a specific research project. The former will be useful for augmenting students' knowledge base of a topic, which will play an important role in presenting ``the big picture," both for original articles and professional presentations. The latter is essential for avoiding duplicating prior work, for determining additional questions to investigate, and for developing a thorough yet concise list of references. For both goals, we describe and demonstrate an algorithm of search, sort, inspect, and repeat.

We assume that students have an initial idea of the appropriate general search terms through discussions with an experienced researcher.
We focus on ISI's Web of Science\cite{isi} because we have found this database to be the most comprehensive and flexible for physics. Other major databases ought to work just as well,\cite{scifind-troy} at least for the physical sciences. In medicine it is necessary to use multiple databases to perform comprehensive searches.\cite{med-search} However, the authors realize that databases such as ISI are usually subscription-only services. Google Scholar\cite{googlescholar} and arXiv.org are free databases that allowing access to a diverse and complementary set of articles, but neither has the flexibility of ISI''s Web of Science. For example, Google Scholar provides a complement to ISI because it searches patents as well as articles. Two current problems with Google Scholar are that its citations are inaccurate, and it does not allow for advanced sorting as we discussed below. The arXiv.org complements ISI because it offers preprints. Two drawbacks to arXiv.org are that it contains un-refereed material, and does not enable advanced sorting.

\section{Tools of the Trade}
We first need to become acquainted with the basic search parameters for the database we have chosen. For ISI these include (a) search type (general, advanced, cited reference), (b) citation database, (c) time range, and (d) ``field tags." Throughout a typical search session, (a)--(c) are fixed, and (d) can change. The most useful combination of these parameters for an active physics researcher is (a) advanced search, (b) SCI-EXPANDED citation database (a filter that eliminates arts and humanities), (c) search all years, and (d) search by topic (\texttt{TS}).\cite{font} For Google Scholar, ``Advanced Scholar Search'' is recommended. To search by topic in the arXiv, use the ``Full Record'' option.

A flexible database is the key to efficient literature searches, and for this reason the following discussion mainly focuses on ISI Web of Science. The most fundamental concepts for efficient searching are field tags, search string perturbations, and ability to sort the results.

\subsection{Field Tags}
Field tags tell the search engine where to look in the database to find entries that compare favorably with the search string.
The most useful of these are \texttt{TS} (topic), \texttt{AU} (author name), \texttt{ZP} (zip/postal code), \texttt{SO} (journal title or source), and \texttt{TI} (article title). These field tags can be combined with boolean operations (\texttt{AND}, \texttt{OR}, \texttt{NOT}) to narrow or expand the search results.
\begin{itemize}
\item \texttt{TS} is the most frequently used field tag and finds matches to the search string in article titles, abstracts, and keywords.

\item \texttt{AU} is useful for perusing a specific researcher's body of work.

\item \texttt{ZP} enables consolidation of search results by postal code, for example, when an author's name is not unique. \cite{footnote1}

\texttt{SO} is used to find articles from a specific journal and is useful when one wants to view articles in the most prestigious journals or a specific field's focus journals.

\texttt{TI} is useful when searching for a single, known article, and sometimes when a \texttt{TS} search yields an unmanageable number of hits. We recommend avoiding \texttt{TI} searches because many articles are missed because the search string was not used in the article's title, though it may be present in the abstract. As an example \texttt{TS=giant magnetoresistance} and \texttt{TI=giant magnetoresistance} have 7402 and 1457 hits, respectively.
\end{itemize}

\subsection{Search Adjustments}
Two important tools for adjusting the search string are phrase searching and truncation. Phrase searching finds exact matches to a string enclosed by quotes (for example, ``\texttt{giant magnetoresistance}''), and is often useful when a specific phrase is common in the field. For example, \texttt{TS=giant magnetoresistance} contains a logical \texttt{AND} between \texttt{giant} and \texttt{magnetoresistance}, but does not distinguish results based on word order. Consequently, spurious results such as ``giant hysteresis of magnetoresistance'' can be avoided by using \texttt{TS}=``\texttt{giant magnetoresistance}.'' To complement this feature, truncation increases the flexibility of searching, and becomes handy when spelling might exclude some hits. Asterisk truncation allows searching for a string that is followed by any number of additional characters. Truncation is most useful when the string might not be in its plural form (for example, \texttt{TS=superlattice*} will find both superlattice and superlattices), or if only the root word is of interest (for example, \texttt{TS=magneto*} will find magnetocaloric, magnetoimpedance, magnetoresistance, magnetosphere). Note that hyphenation is not an issue for \texttt{TS}, at least with magnetoresistance, which is sometimes spelled magneto-resistance. In cases where one character may or may not be present, as is commonly a problem with British and American spellings, a \$ placed in the location of the optional character will yield all hits with and without that character: \texttt{TS=quark flavo\$r} is equivalent to \texttt{TS=(quark AND (flavor OR flavour))}.

Refining and iterating the search algorithm is key. But how do students learn to meaningfully refine their searches? Initially, a more senior researcher will be necessary to guide the student in this respect. It is also important to know a research project's boundary conditions (for example, specific instrumentation may be desirable). Additional perturbations can be discovered by identifying keywords or concepts that appear repeatedly in a subset of the articles. Searching the literature is a skill, and students need to realize that ``practice makes perfect."

\subsection{Sorting}
How a particular search engine orders results is a key feature of that database. The options available for sorting results vary by database. When a search is performed in ISI Web of Science, the results can be sorted by a number of parameters. Most significantly, results can be sorted by date or by number of citations. Google Scholar automatically orders the results using a complicated algorithm that considers factors such as the date, times cited, author, and journal. There is an option to find ``recent articles'' in Google Scholar that will increase the importance of the date in this algorithm. There are no other ways to customize the way that Google Scholar sorts results. The results of a search using the arXiv are always sorted by date, without any options for customization.

Sorting by the number of citations is important when trying to locate the articles which are most highly valued in a particular field. By examining the most cited articles in a search result, the key ideas and necessary background information for a topic can be easily obtained.
This method does not imply that articles with low citations are necessarily of lesser importance, as is obvious if we consider recent publications that have not had time to be judged by the scientific community. Furthermore, it is possible that a highly refuted article makes it into the top-cited list purely because it purports ideas that are unpopular.

Sorting by date can also be useful to determine the timeliness of the results and to examine those articles that have not yet had a chance to receive citations. ISI will return the latest articles published in peer-reviewed journals. Often, even more recent results are desired, and the arXiv is an invaluable resource for this type of searching, and is the best way to find preprints and relevant articles about ongoing studies. However, because the results returned from the arXiv are not peer-reviewed, searching the arXiv alone is not a sufficient way to research a topic.

\subsection{Advanced Sorting: The $h$-index}

\begin{figure}[h]
\centering
\includegraphics[width=3.in]{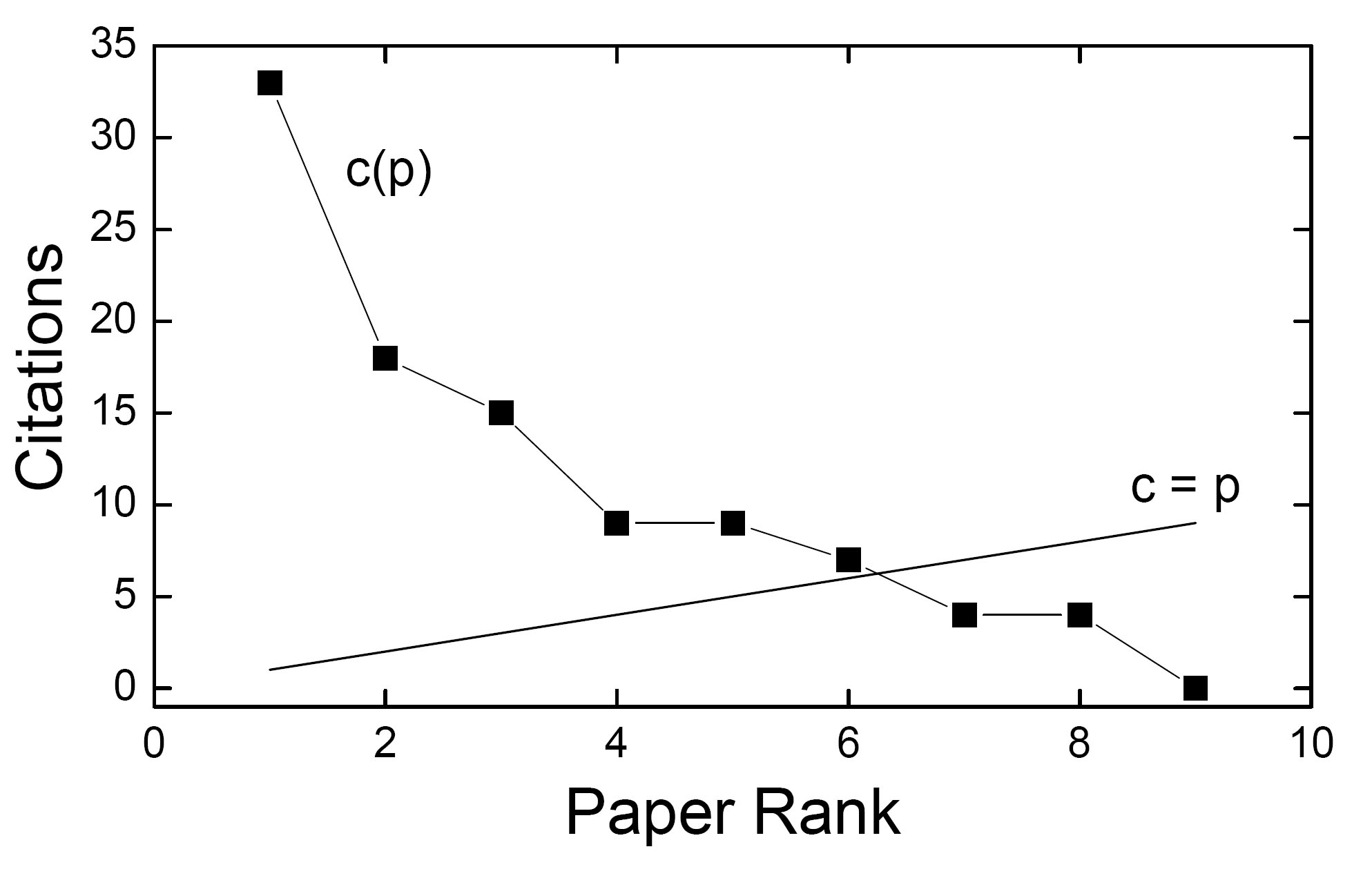}
\caption{Example graphical determination of the $h$-index for a collection of articles with $h=6$. The rank nearest to the intersection of the line $c = p$ with $c(p)$ indicates $h$.}
\label{hindex}
\end{figure}

A useful way of sorting a search is by number of citations, which allows inspection using the $h$-index concept.\cite{hfactor} The $h$-index is determined as follows. The articles are first ranked by the number of times that they have been cited such that the article ranked 1 has the most citations.  Thus, since the citations descend as rank ascends, the rank must exceed the number of citations somewhere on the list; the rank where the crossover occurs is defined as the $h$-index.  As illustrated in Tab.~\ref{htable}, a set of articles with $h=6$ has six articles each with six or more citations. The seventh article in this example has a rank greater than its citations. It may be instructive to see how $h$ can be determined graphically: $h$ is the rank nearest to the intersection of $c(p)$ with the line $c=p$, where $c(p)$ is the number of citations for paper $p$ in the ordered list (Fig.~\ref{hindex}).

Applying the $h$-index to individuals has proven to be very
effective. An individual's $h$-index is found by ranking that individual's articles by times cited, and finding the value at which the rank equals the number of citations. Hirsch showed that successful scientists have large $h$ indices (and more importantly a large value of $dh/dt$).\cite{hfactor} Although Hirsch presented considerable evidence suggesting that no universal criterion exists for determining whether or not an absolute $h$-index can be qualified as ``good" or not, larger values of $h$ always implies ``more influential." A hind-sight study showed a strong correlation between $h$ and committee peer review,\cite{doeshwork} and indicates that $h$ measures how one's contributions are viewed by one's peers. Without loss of generality, we may apply the $h$-index to any collection of articles returned from a search to find a subset of the most influential (most highly cited) articles correlated with the search string.\cite{Banks-h} This use of $h$ can be a useful guideline to determine approximately how many articles returned in a search should be examined. For example, we could read the top $h$ articles to sample the most influential work in a specific area. As mentioned, this method does not imply that articles with rank greater than $h$ are of poor quality.

\section{Example Searches}
The basic methodology we employ is search, sort, inspect, and repeat. Searching is performed with the field tags and search string perturbations we have introduced. Sorting is performed to determine the highest ranked articles of a specific search. Inspecting the top-cited articles of a search will lead us to the most influential topics or keywords within the results, which will then allow us to refine the search string, and then repeat these steps.

\begin{figure}[h]
\centering
\includegraphics[width=6.25in]{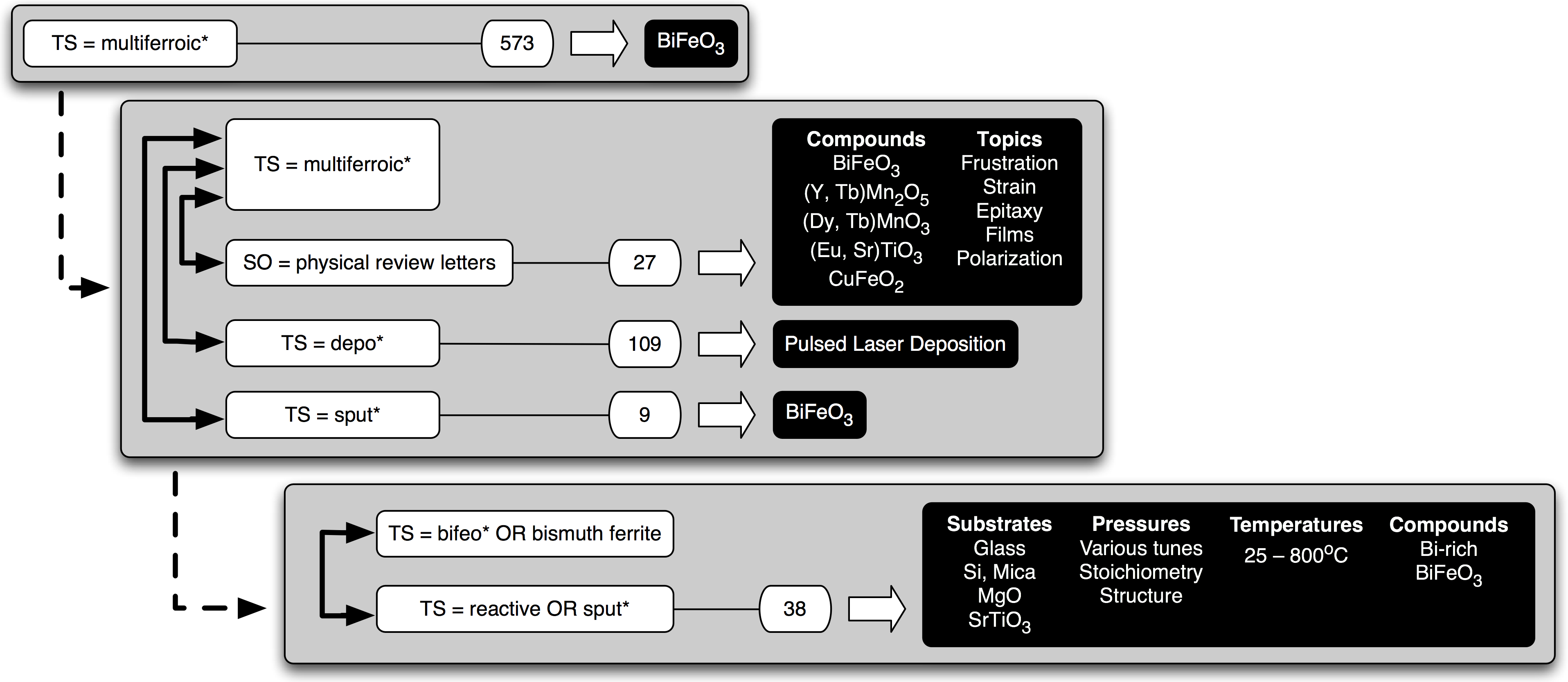}
\caption{Mapping the search. Search strings are in white boxes.  Double-headed arrows indicate a logical AND.  Number of hits from a search are connected with thin lines. Block arrows lead to conclusions (collected in black boxes) drawn by inspecting articles returned from the different branches of the search.  Dashed arrows indicate a transition from general to more specific levels of searching.}
\label{map}
\end{figure}

We will use this methodology to perform two example searches on the topic of multiferroics, one general and one specific. For the latter, we imagine a student having been asked to determine if, what, and how multiferroics have been grown by sputter deposition. A set of articles uncovered by various searches allows us to draw several conclusions about the field in general, and the deposition process of one specific compound. A summary of both the general and specific searches is presented as a flow diagram in Fig.~\ref{map}, including search strings, number of hits per search, and conclusions drawn from the article sets returned by the database.\cite{old}

The efficiency of this method relies on active researchers' interpretations of previously published work in their field -- their citations are an indication of an article's quality. The information we learn about multiferroics gives us a general idea of the important issues in this area without having to read more than titles. More detailed information can be obtained by reading abstracts to identify recurring ideas or topics. Ultimately we will have to read the articles in their entirety to judge their appropriateness.

\subsection{General Search} 
\begin{figure}[h!]
\centering
\includegraphics[width=3.in]{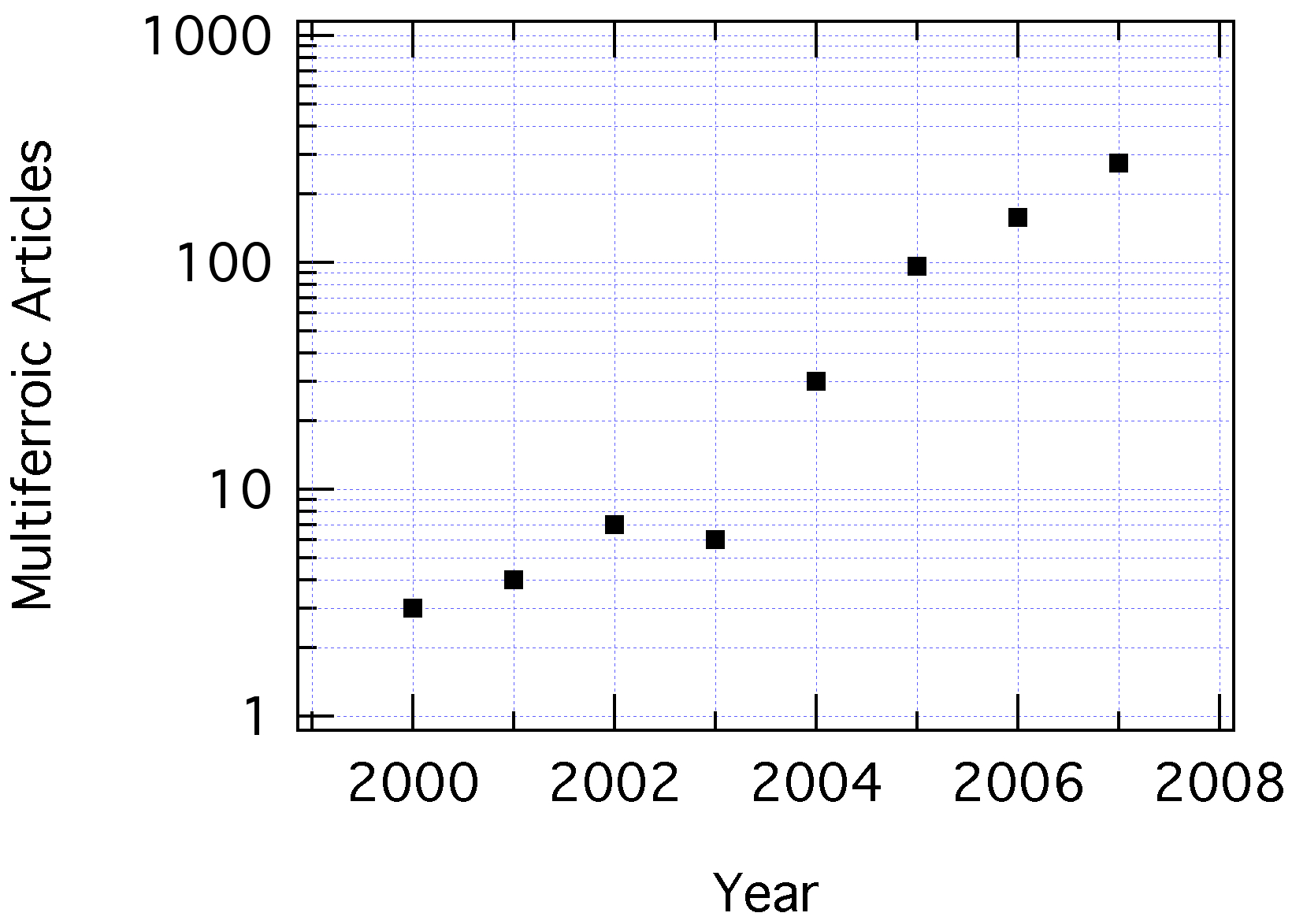} 
\caption{A plot of the annual number of multiferroic publications versus publication year shows that this topic is just starting, and interest in it is increasing rapidly. }
\label{yearly-pubs}
\end{figure}

We begin our general search with \texttt{TS=multiferroic}, which yields 504 hits -- a lot to read, but less so if 50 hits per page are shown. For completeness, we then adjust to \texttt{TS=multifer*} and find 630. If we sort by times cited, we find that the some of the top articles contain the word ``multifermentans," which appears in biochemistry journals. These are errant hits, so we adjust the search to \texttt{TS=multiferr*} and obtain 586 hits. Of the top-cited articles, the majority are in respectable journals: \PRB, \PRL, \textit{Science}, \textit{Nature}, \textit{Nature Materials}, \APL, and \JAP. We notice by perusing the titles that many articles are concerned with bismuth ferrite, BiFeO$_3$. We also notice that most of the top-cited articles have been published within the last five years. We can extract the number of articles published each year, plot the results as shown in Fig.~\ref{yearly-pubs} and see that this field is really just starting to take off. (To do this, go to ``Analyze Results" and rank the records by publication year, or use the ``Create Citation Report" option.) In so doing, we notice that one anomalous result was published in 1996, several years before the majority of multiferroic articles. This article is entitled ``Clusters of lymphoma in ferrets," and should clearly be disregarded in the analysis. Near this article, we discover another errant hit published in the \textit{Journal of Coordination Chemistry}, which passed through the filter because it is concerned with ``multiferrocenyl groups." At this point, we refine our search systematically and use \texttt{multiferro*} (584 hits), \texttt{multiferroi*} (576 hits), \texttt{multiferroic*}(573 hits). With 	
\texttt{(TS=multiferro*) NOT TS=(multiferroic*)}, we realize that \texttt{multiferrocenyl} accounts for the majority of the eleven article difference between these two search strings. Ultimately, we choose \texttt{multiferroic*}, which takes care of the plural and singular usages. Alternatively, we could have searched \texttt{(TS=multiferro* NOT multiferrocenyl)} to obtain the relevant 573 articles. Having convinced ourselves that this search string contains the most appropriate articles, we now proceed to investigate either the big picture or specific details related to multiferroics.

To get the ``big picture,'' we chose to view articles published in the top journals. \textit{Physical Review Letters} is one of the most prestigious physics journals, so viewing its articles will give us a glimpse of the top research on multiferroics.\cite{footnote2} This search is achieved with the string \texttt{TS=multiferroic*} \texttt{and} \texttt{SO=Physical} \texttt{Review Letters}. There are only 27 articles that meet these criteria, making this search an excellent starting point. By sorting by times cited, we notice a variety of compounds in addition to the previously identified BiFeO$_3$: YMn$_2$O$_5$, TbMn$_2$O$_5$, DyMnO$_3$, EuTiO$_3$, SrTiO$_3$, TbMnO$_3$, and CuFeO$_2$. In this manner we have discovered that multiferroics research focuses on compounds containing oxides of Fe, Mn, and Ti. It is likely there are other multiferroic compounds being studied, but this procedure has identified those with the most success at the present time. We can learn more by quickly scanning the titles of these articles for recurring words or concepts. This scan makes it apparent that the following topics are important: frustration, strain, epitaxy, films, and polarization.

For a deeper understanding of current issues we should read the introductory paragraphs of the articles published in \PRL. Because \PRL\ is intended to have a broad readership, the introductions are more accessible to someone just learning about the field (the body of the article may be another story). Further, because the aim of a good introduction is to describe the background and history of the work presented in the article,\cite{goodintro} any work cited in the introduction should also be read. After reading a dozen or so articles, paying particular attention to their introductions and conclusions, we will begin to uncover the big picture.

\subsection{Specific Search}
For specific details we must know more or less what we are looking for, but should begin using broad terms. Starting a search that is too focused will inevitably exclude pertinent articles because different authors choose slightly different words or methods to describe or investigate the same problem. Suppose, for instance, that we are interested in sputter deposition of multiferroics. We should first determine if sputtering has ever been used to grow any multiferroic films, then aim to determine specific materials and deposition parameters (temperature, pressure, or substrate type). \texttt{TS=multiferroic*} \texttt{sputtering} yields 7 hits; \texttt{TS=multiferroic*} \texttt{sputter*} yields 9 hits; \texttt{TS=multiferroic*} \texttt{sput*} yields 9 hits. Now we should be convinced that very few of the 573 multiferroic articles have sputtering as the thin film deposition technique. What method is preferred? A search using \texttt{TS=multiferroic* depo*} yields 109 hits. Sorting by times cited, this search has an $h$ of 14. The majority of the top-cited articles use pulsed laser deposition, and only one used sputtering. Because we are concerned with sputtering, we return to the 9 articles on that topic as a starting point. From these titles we discover that BiFeO$_3$ is the most commonly grown material by (reactive) sputtering. We therefore start a new search: \texttt{TS=((bifeo*} \texttt{OR bismuth} \texttt{ferrite)} \texttt{AND} \texttt{(reactive} \texttt{OR sput*))} with 38 hits, several of which are in \APL\ and \JAP, which are more specialized journals than \PRL. A quick glance at the abstracts reveals that this material has been grown on glass, mica, Si, MgO, and SrTiO$_3$ (using various buffer layers) at temperatures from ambient to $800^{\circ}$C, and pressure is used to tune the stoichiometry and structure. Further, the films are reactively sputtered from off-stoichiometric (Bi-rich) compound targets.

\subsection{Putting it all together}
By using the tools and methodology we have presented, we have uncovered general multiferroic references, and specific references related to sputter deposition of bismuth ferrite (see Fig.~\ref{map}). The general references (and pertinent references contained therein) will assist us in understanding the global issues in multiferroics. The deposition references (and relevant references contained therein) will help determine where we might begin to learn how to grow BiFeO$_3$ by sputtering, what obstacles we might encounter, and what has already been done with this technique. Following a trail of references will lead to additional general and specific information. Usually one to two degrees of separation is sufficient to find closely related articles. For a complete picture that surrounds a specific article, we must inspect both its cited and citing articles.

\section{Student Perspectives}

We have implemented the search procedure we have described in undergraduate and graduate courses. Most graduate students found the methodology to be straightforward, citing previous search experience. They particularly appreciated the field tags and search adjustments, and several remarked that this methodology helped them find articles related to their research projects that they had not previously found.  In what follows, reported errors are estimated by standard deviation of the mean.

\begin{figure}[h!]
\centering
 \includegraphics[width=3.2in]{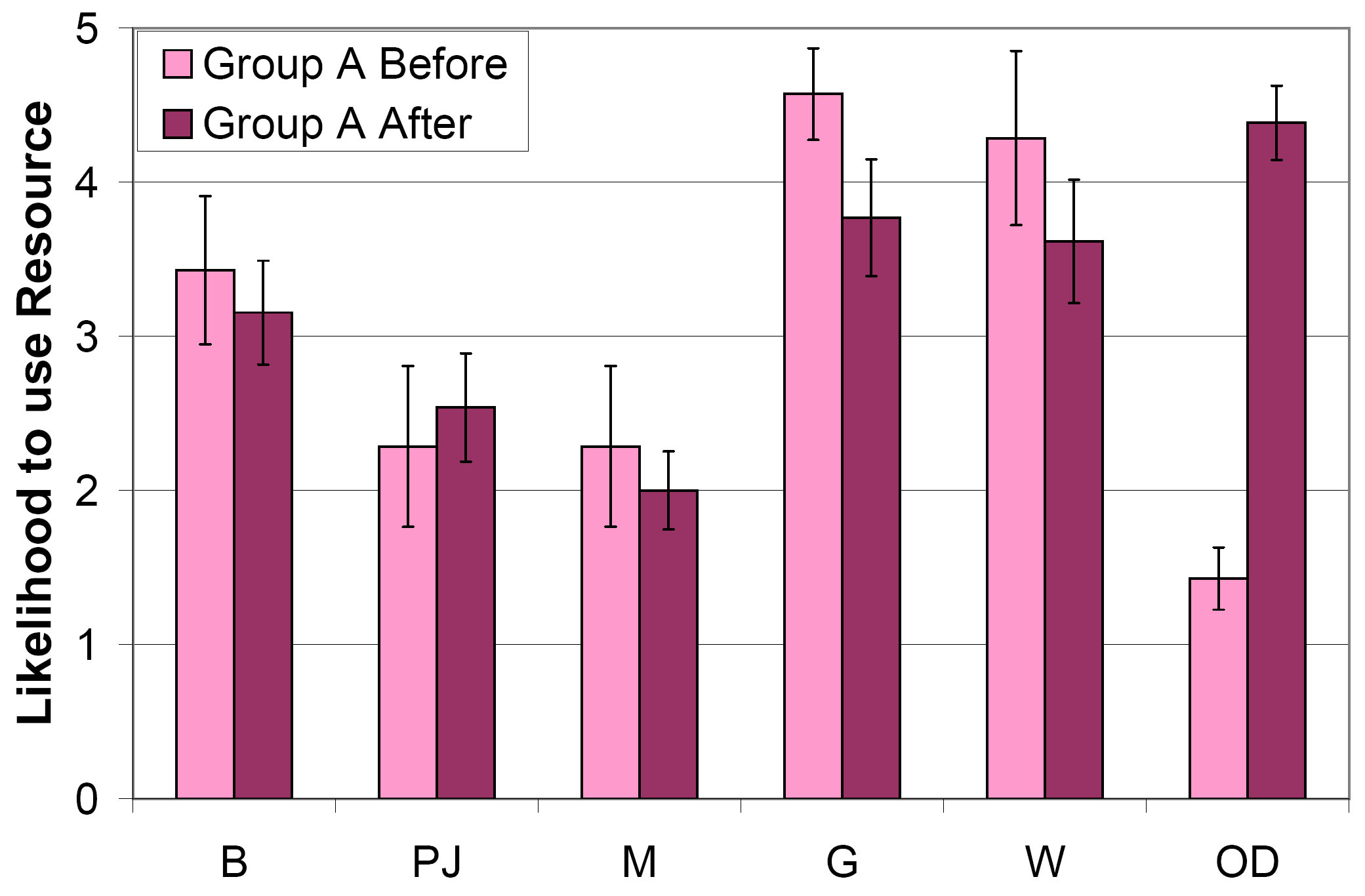}\\
\includegraphics[width=3.2in]{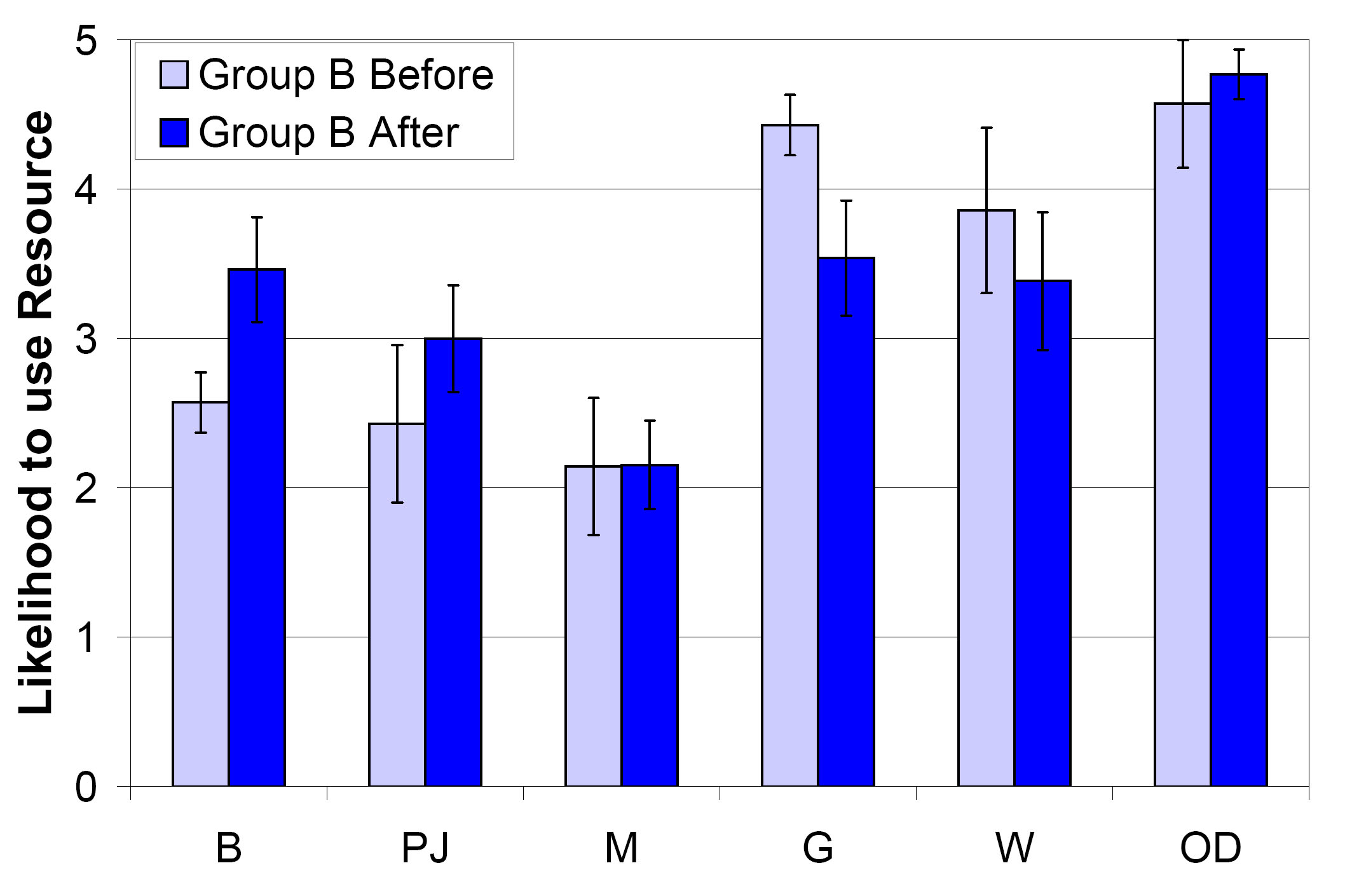}
\caption{(Color online) Likelihood of Groups A and B to use available resources before and after our methodology was introduced, based on a 1--5 scale with 1 = ``never" and 5 = ``always." The categories were B = books, PJ= print journals, M = magazines, G = Google, W = Wikipedia, and OD = online journal databases. }
\label{survey}
\end{figure}

\begin{figure}[h!]
\centering
\includegraphics[width=3.2in]{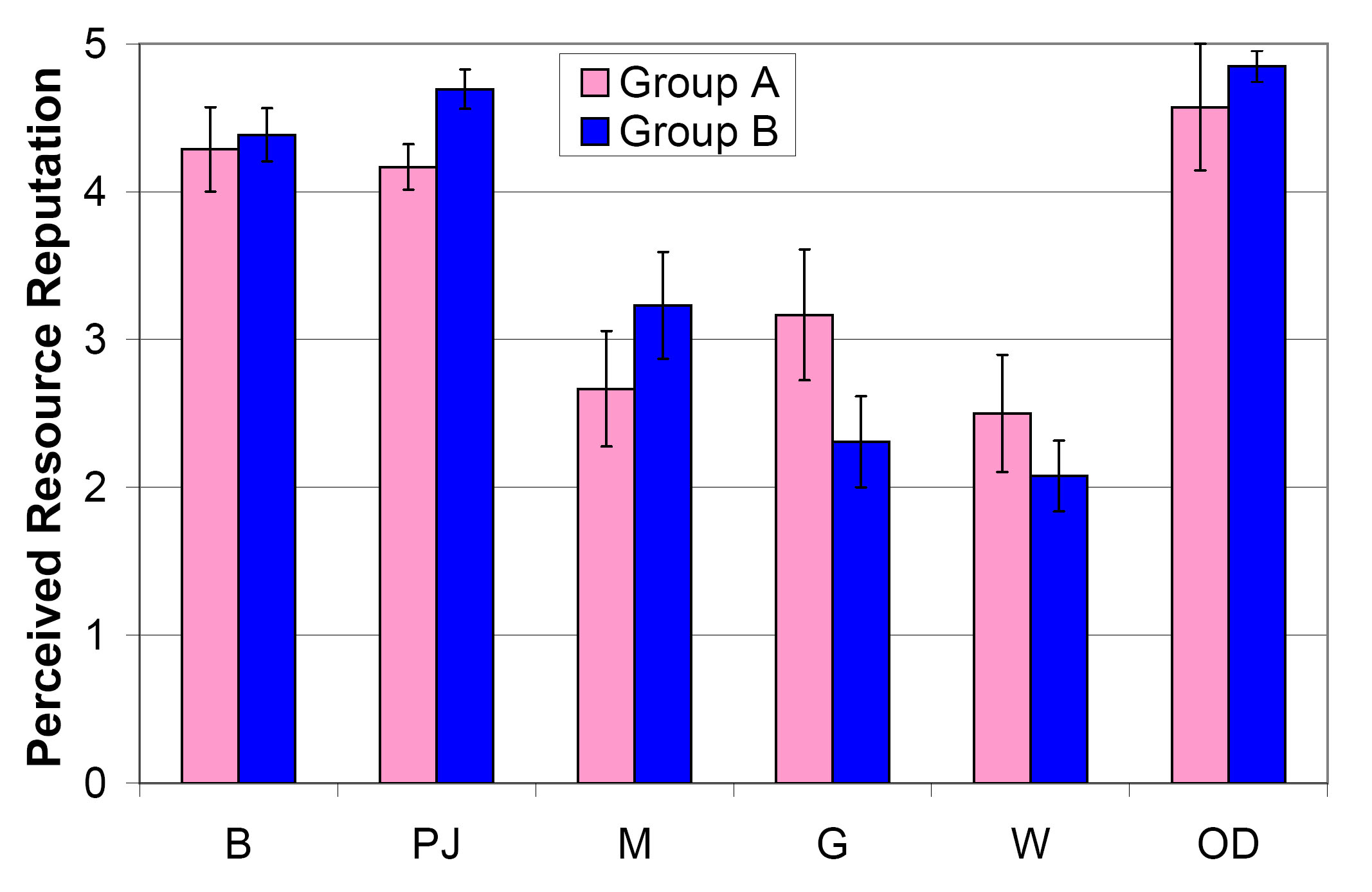}
\caption{(Color online)
Students' perceptions of the scientific reputation of available resources on a scale of 1--5 with 1 = ``least reputable" and 5 = ``most reputable." The categories were B = books, PJ= print journals, M = magazines, G = Google, W = Wikipedia, and OD = online journal databases. }
\label{rep}
\end{figure}

Twenty undergraduate ``Advanced Laboratory" students were surveyed to determine their impressions of the methodology, as well as to quantify its impact. On a scale of 1--5 with 5 being the best, the total perceived usefulness of this methodology was $4.4\pm0.2$, with 60\% responding ``5." Based on their indications of prior online journal database use, 9 students were identified to have ``little or no prior use" (Group A), and 11 were identified as having ``some prior use" (Group B). To gauge the impact of this experience, they were asked to indicate what resources they would have used to gather information for science term papers prior to and after learning this methodology (Fig.~\ref{survey}). The options included books, print journals, magazines, Google, Wikipedia, and online journal databases; the scale was 1--5, with 1 indicating ``never,'' and 5 ``always." Group A showed a statistically significant change in the use of online journal databases, which jumped from $1.4\pm0.2$ to $4.4\pm0.2$. Of marginal statistical significance for Group A was a decreased mean and increased variance for future use of Google. Of marginal statistical significance for Group B was an increased mean and increased variance for future use of books, and a decreased mean and increased variance of future use of Google. The increased variance implies that students' perception of Google as an appropriate resource decreased. This sentiment is echoed in Fig.~\ref{rep}, which shows the students' perception of where to find reputable science. Group A holds Google and Wikipedia in higher esteem than Group B, and Group B holds books, journals, and magazines in higher esteem than Group A. It is apparent that Group B is more closely reflecting the responses of professional scientists, which suggests that repeated exposure is needed to solidify students' perception of where one is most likely to find reputable scientific results.  To this end, we note the significantly lower variance of Group B's rankings of online journal databases. Thus, this endeavor has provided a solid first step for new users, as indicated by the very large increase in likelihood to use online journal databases, and has also helped to reinforce which scientific resources are reputable among more experienced students.

\section{Summary}
Knowing how to use an online database for literature searches can lead to a significant amount of information in a relatively short period of time if performed in a systematic fashion. The method described here allows one to converge on the desired information in a limited number of searches, whether that information is of a general or specific nature, by using a database that allows sorting by the number of citations. The techniques demonstrated here will help students of any level find influential articles related to a given topic, be it for a term paper or a research project. Students involved in this study showed marked improvement in their awareness of where to find sound scientific information.

\begin{acknowledgments}
Work at USF was supported by the National Science Foundation.
Work at Centenary College was supported by the Mattie Allen Broyle's Inaugural Year Research Chair and the Gus S. Wortham Endowed Chair of Engineering.
\end{acknowledgments}

\newpage\section*{Figure captions}

\begin{table}[h]
\caption{Example determination of the $h$-index for a collection of articles with $h=6$. The double line indicates the boundary defining $h$.}
\label{htable}
\begin{center}
\begin{tabular}{ccc}
\hline \textbf{~~~~Rank~~~~}&  ~~&    \textbf{~~Times Cited~~}\\\hline
1 & $<$ & 33 \\
2 & $<$ & 18 \\
3 & $<$ & 15 \\
4 & $<$ & 9 \\
5 & $<$ & 9 \\
6 & $\le$ & 7 \\
\hline\hline
7 & $>$ & 5 \\
8 & $>$ & 5 \\
9 & $>$ & 0 \\
\hline
\end{tabular}
\end{center}
\end{table}

\begin{thebibliography}{11}
\expandafter\ifx\csname natexlab\endcsname\relax\def\natexlab#1{#1}\fi
\expandafter\ifx\csname bibnamefont\endcsname\relax
\def\bibnamefont#1{#1}\fi
\expandafter\ifx\csname bibfnamefont\endcsname\relax
\def\bibfnamefont#1{#1}\fi
\expandafter\ifx\csname citenamefont\endcsname\relax
\def\citenamefont#1{#1}\fi
\expandafter\ifx\csname url\endcsname\relax
\def\url#1{\texttt{#1}}\fi
\expandafter\ifx\csname urlprefix\endcsname\relax\def\urlprefix{URL }\fi
\providecommand{\bibinfo}[2]{#2}
\providecommand{\eprint}[2][]{\url{#2}}

\bibitem[{\citenamefont{of~California~Berkeley}(2007)}]{UCBtutorial}
\bibinfo{author}{\bibfnamefont{University of California Berkeley}},
\bibinfo{title}{``Teaching library internet workshops"}
, \bibinfo{note}{\url{<www.lib.berkeley.edu/TeachingLib/ Guides/ Internet/ FindInfo.html>}}.

\bibitem[{\citenamefont{Hirst}(2005)}]{Manchestertutorial}
\bibinfo{author}{\bibfnamefont{D.}~\bibnamefont{Hirst}}, \bibinfo{title}{``Conducting a literature search,"} \bibinfo{year} \bibinfo{note}{\url{< www.mace.manchester.ac.uk/ aboutus/ informationcentre/ informationskills/ searchguides/guideforstudents.pdf>}}.

\bibitem[{\citenamefont{Espinoza et~al.}(2006)\citenamefont{Espinoza, Rincon, and Chacin}}]{healthsci}
\bibinfo{author}{\bibfnamefont{N.}~\bibnamefont{Espinoza}}, \bibinfo{author}{\bibfnamefont{A.}~\bibnamefont{Rincon}}, \bibnamefont{and} \bibinfo{author}{\bibfnamefont{B.}~\bibnamefont{Chacin}}, \bibinfo{title}{``Use of web-based tools by health sciences professionals at a Venezuelan university for searching scientific literature. A cross-sectional study,"} \bibinfo{journal}{Profesional de La Informaci\'{o}n} \textbf{\bibinfo{volume}{15}}, \bibinfo{pages}{28--33} (\bibinfo{year}{2006}).

\bibitem{isi}\url{<www.isiknowledge.com>}.

\bibitem{scifind-troy}
\bibinfo{note}{For a supplemental search of the Chemical Abstracts Service (CAS) database via SciFinder Scholar, see \url{<www.centenary.edu/attachments/physics/tmessina/litsearch/literaturesearchsi01.pdf>}}.

\bibitem[{\citenamefont{Smith et~al.}(2004)\citenamefont{Smith, Herzka, Wenz, and Henze}}]{med-search}
\bibinfo{author}{\bibfnamefont{C.~G.} \bibnamefont{Smith}}, \bibinfo{author}{\bibfnamefont{A.~S.} \bibnamefont{Herzka}}, \bibinfo{author}{\bibfnamefont{J.~F.} \bibnamefont{Wenz}}, \bibnamefont{and} \bibinfo{author}{\bibfnamefont{E.~P.} \bibnamefont{Henze}}, \bibinfo{title}{``Searching the medical literature,"} \bibinfo{journal}{Clinical Orthopaedics and Related Research} \textbf{\bibinfo{volume}{421}}, \bibinfo{pages}{43--49} (\bibinfo{year}{2004}).

\bibitem[{goo()}]{googlescholar}
\url{<www.scholar.google.com>}.

\bibitem{font}Text appearing in typewriter font indicates search strings used in online databases.

\bibitem{footnote1}Careful attention must be paid to work done at different stages of an author's career, because academics often change institutions.

\bibitem[{\citenamefont{Hirsch}(2005)}]{hfactor}
\bibinfo{author}{\bibfnamefont{J.~E.} \bibnamefont{Hirsch}}, \bibinfo{title}{``An index to quantify an individual's scientific research output,"} \bibinfo{journal}{Proc. Nat. Acad. Sci. USA} \textbf{\bibinfo{volume}{102}}, \bibinfo{pages}{16569--16572} (\bibinfo{year}{2005}).

\bibitem[{\citenamefont{Bornmann and Daniel}(2005)}]{doeshwork}
\bibinfo{author}{\bibfnamefont{L.}~\bibnamefont{Bornmann}} \bibnamefont{and} \bibinfo{author}{\bibfnamefont{H.-D.} \bibnamefont{Daniel}}, \bibinfo{title}{``Does the h-index for ranking of scientists really work?,"} \bibinfo{journal}{Scientometrics} \textbf{\bibinfo{volume}{65}}, \bibinfo{pages}{391--392 } (\bibinfo{year}{2005}).

\bibitem[{\citenamefont{Banks}(2006)}]{Banks-h}
\bibinfo{author}{\bibfnamefont{M.~G.} \bibnamefont{Banks}}, \bibinfo{title}{``An extension of the Hirsch index: Indexing scientific topics and compounds,"} \bibinfo{journal}{Scientometrics} \textbf{\bibinfo{volume}{69}}, \bibinfo{pages}{161--168} (\bibinfo{year}{2006}).

\bibitem{footnote2} If the prestige of a journal is unknown, we can estimate the journal's overall impact by analyzing its h-index in analogy to the h-index of an individual.

\bibitem{old}All search results were found in late December, 2007, and will have changed by publication of this article.

\bibitem{goodintro}
Editorial, ``The aim of a good introduction," \url{<prl.aps.org/ edannounce/ PRLv95i17.html>}.

\end{thebibliography}
\end{document}